\documentclass[a4paper]{article}

\usepackage{INTERSPEECH2021}
\usepackage{graphicx}
\usepackage{subfigure}
\usepackage{multirow}
\usepackage{bm}
\usepackage{threeparttable}

\title{One in A Hundred: Selecting the Best Predicted Sequence from Numerous Candidates for Streaming Speech Recognition}

\name{Zhengkun Tian$^{1,2}$, Jiangyan Yi$^{1}$, Ye Bai$^{1,2}$, Jianhua Tao$^{1,2,3}$, Shuai Zhang$^{1,2}$, Zhengqi Wen$^{1}$}
\address{
  $^1$National Laboratory of Pattern Recognition, Institute of Automation, \\ Chinese Academy of Sciences, Beijing, China \\
  $^2$School of Artificial Intelligence, University of Chinese Academy of Sciences, Beijing, China\\
  $^3$CAS Center for Excellence in Brain Science and Intelligence Technology, Beijing, China}
\email{\{zhengkun.tian, jiangyan.yi, jhtao, ye.bai, shuai.zhang, zqwen\}@nlpr.ia.ac.cn}

\begin{document}

\maketitle
\begin{abstract}
The RNN-Transducers and improved attention-based encoder-decoder models are widely applied to streaming speech recognition. Compared with these two end-to-end models, the CTC model is more efficient in training and inference. However, it cannot capture the linguistic dependencies between the output tokens. Inspired by the success of two-pass end-to-end models, we introduce a transformer decoder and the two-stage inference method into the streaming CTC model. During inference, the CTC decoder first generates many candidates in a streaming fashion. Then the transformer decoder selects the best candidate based on the corresponding acoustic encoded states. The second-stage transformer decoder can be regarded as a conditional language model. We assume that a large enough number and enough diversity of candidates generated in the first stage can compensate the CTC model for the lack of language modeling ability. All the experiments are conducted on a Chinese Mandarin dataset AISHELL-1. The results show that our proposed model can implement streaming decoding in a fast and straightforward way. Our model can achieve up to a 20\% reduction in the character error rate than the baseline CTC model. In addition, our model can also perform non-streaming inference with only a little performance degradation.

\end{abstract}
\noindent\textbf{Index Terms}: Streaming Speech Recognition, End-to-End Models, Hybrid CTC and Attention, Two-Stage Inference
\vspace{-10pt}
\section{Introduction}

Streaming speech recognition has been applied to many real scenarios, like meeting real-time transcription and keyboard dictation systems on mobile phones.  The mainstream models for streaming speech recognition include RNN-Transducer (RNN-T) models and the improved attention-based encoder-decoder models \cite{jaitly2015neural,sainath2018improving,he2018streaming,moritz2019triggered,dong2020cif,Tian2019,tian2020synchronous}.

The RNN-T model, which utilizes unidirectional recurrent neural networks as the encoder, can be directly applied to streaming speech recognition \cite{he2018streaming,Tian2019,graves2012sequence,graves2013speech}. However, the RNN-T model suffers from the inefficiency of training and inference \cite{bagby2018efficient, li2019improving}. The traditional attention-based encoder-decoder model decodes the output sequences based on the previously predicted tokens and the entire acoustic encoded states, which prevents it from decoding the output sequence in a streaming way \cite{tian2020synchronous}. The improved attention-based encoder-decoder models mainly include the following versions: the monotonic chunk-wise attention (MoChA) \cite{chiu2017monotonic}, the triggered attention \cite{moritz2019triggered}, the continuous integrate-and-fire (CIF) \cite{dong2019cif}, the synchronous transformer \cite{tian2020synchronous}, and so on. These models utilize many complicated methods to segment the acoustic encoded states and then compute the attention weights on the segment states. As a kind of end-to-end model, connectionist temporal classification (CTC) models were first proposed to transcribe the acoustic feature sequences into the corresponding text sequence \cite{graves2006connectionist,hannun2014deep,amodei2016deep}. The CTC models cannot capture the linguistic dependencies between the output tokens \cite{graves2012sequence}. To address this problem, CTC models tend to depend on an external language model or a WFST-based graph to improve the performance \cite{amodei2016deep, miao2015eesen}. Recently, some very deep convolution CTC models, like Jasper \cite{li2019jasper} and ContextNet \cite{han2020contextnet}, have achieved competitive performance with other end-to-end models. Although very deep convolution layers make the model able to model long-range contexts, it will also result in a large latency and prevent the model from being applied to streaming speech recognition.

The CTC models are more efficient in inference than the other two end-to-end models. Inspired by the success of two-pass end-to-end models \cite{sainath2019two}, we introduce a transformer decoder and the two-stage inference method into the CTC model to improve the model performance on streaming speech recognition. Now that the CTC model cannot model the linguistic dependencies between the output tokens, the transformer decoder can be regarded as a conditional language model, which can pick out the best sequence from the top-N candidates based on the linguistic knowledge and the corresponding acoustic information. We assume that a large enough number and diversity of candidates generated by the CTC model can compensate for the lack of language modeling ability. We rename the process that selecting the best sequence from numerous candidates as On-In-A-Hundred(OAH). Our model consists of three components, a latency-controlled streaming transformer encoder, a CTC decoder, and a transformer decoder. We first force the self-attention mechanism in the transformer encoder to focus on the local context to model the input sequence in a streaming fashion. Then we introduce a latency-controlled context layer at the top of the transformer encoder to model the future context of limited range. During training, these three components are optimized jointly. The one-in-a-hundred inference process can be divided into two stages: pre-selection and one-step scoring. At the pre-selection stage, the CTC decoder generates many possible sequences as candidates. At the one-step scoring stage, the transformer decoder scores all these candidates based on the corresponding acoustic encoded states and then selects the sequence with the highest average score as the final predict sequence. All experiments are conducted on a public Chinese Mandarin dataset AISHELL-1. The results show that our proposed model can achieve up to a 20\% reduction in the character error rate compared to the baseline CTC model. Furthermore, our model can also perform non-streaming decoding.

\begin{figure*}[t]
	\centering
	\subfigure[Model Architecture]{
		\centering
		\label{fig:model}
		\includegraphics[width=0.3\linewidth]{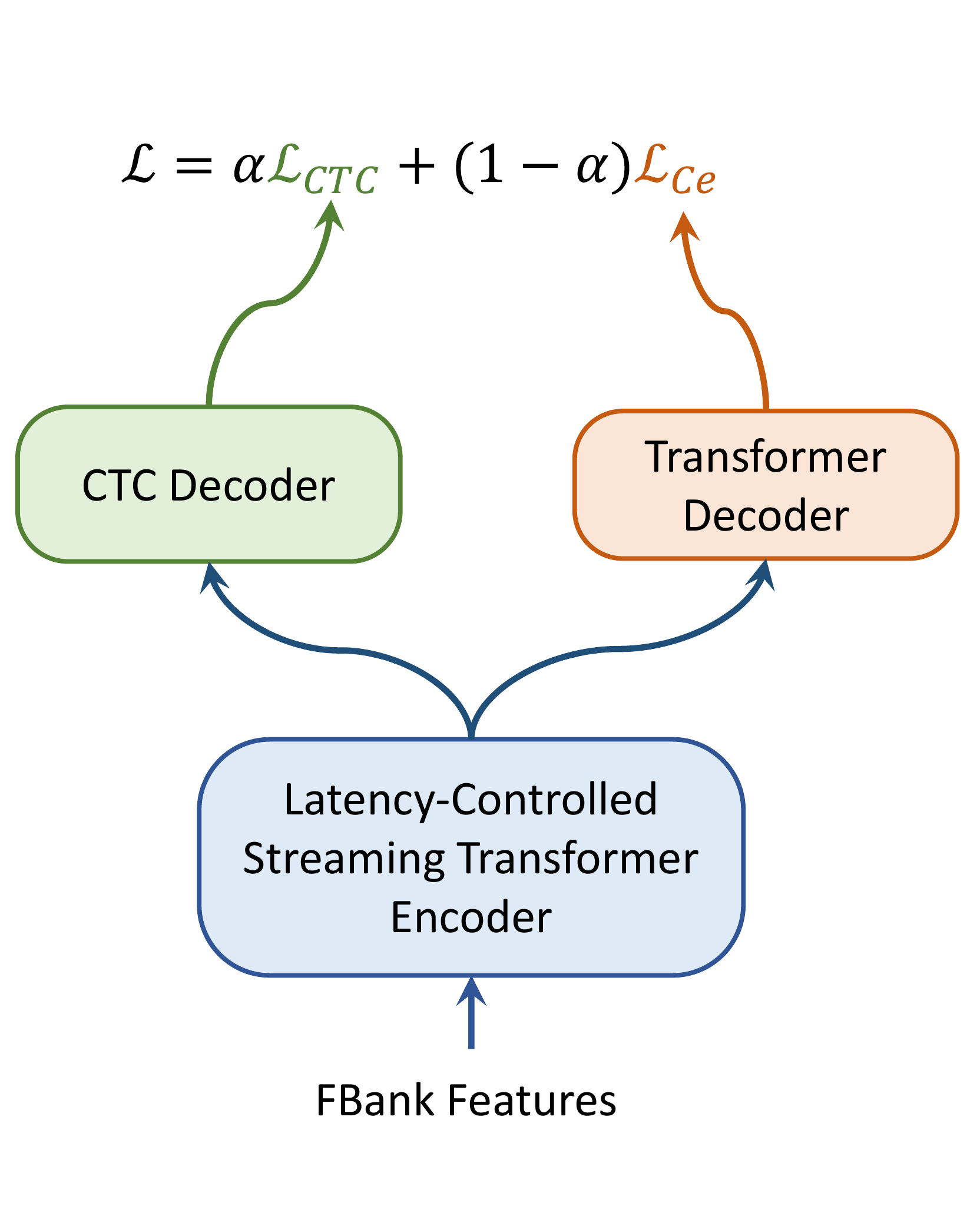}}
	\subfigure[Encoder Architecture]{
		\centering
		\label{fig:encoder}
		\includegraphics[width=0.25\linewidth]{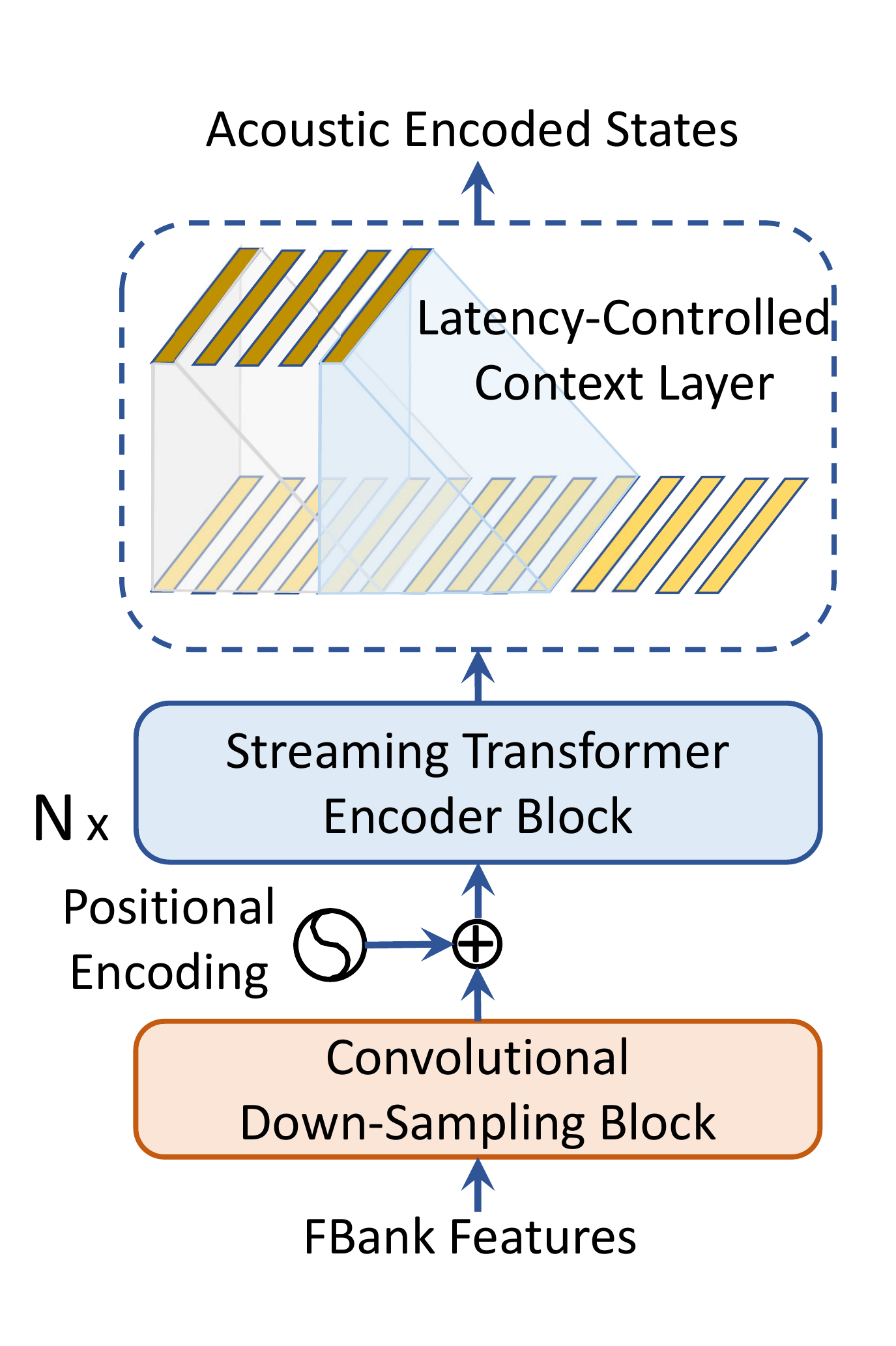}}
	\subfigure[One-In-A-Hundred Inference]{
		\centering
		\label{fig:inference}
		\includegraphics[width=0.4\linewidth]{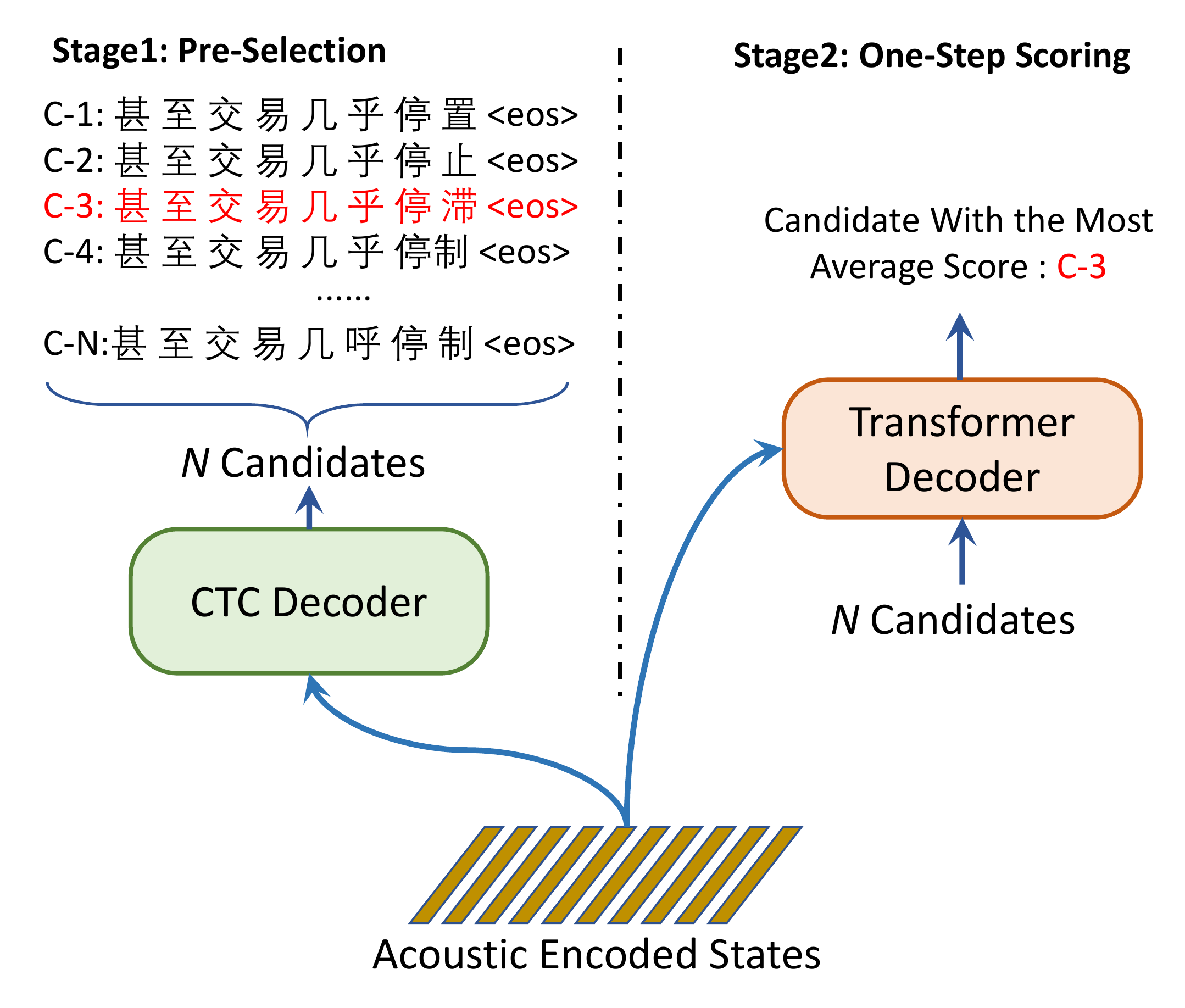}}
	\vspace{-15pt}
	\caption{(a) illustrates the overall structure of our proposed model. The model consists of a latency-controlled streaming transformer encoder (LC-STE), a CTC decoder, and a transformer decoder. (b) illustrates the details of the encoder. The LC-STE consists of a convolutional down-sampling block, N streaming transformer blocks, and a latency-controlled context layer. (c) illustrates the one-in-a-hundred inference process. The inference process can be split into two stages: pre-selection and one-step scoring. The CTC decoder can generate many possible candidates at the pre-selection stage. Then, the transformer decoder selects the best predicted sequence from the candidates in one step. The best candidate C-3 means that "the deal is almost stalled".}
	\vspace{-20pt}
\end{figure*}

The remainder of this paper is organized as follows. We introduce the details of the model and related works in Section 2 and Section 3 respectively. Our experimental setup and results will be presented in Section 4. The conclusions will be given in Section 5.
\vspace{-5pt}
\section{Our Proposed Method}
\vspace{-5pt}
Our model consists of three components, a latency-controlled streaming transformer encoder, a CTC decoder, and a transformer decoder, as shown in Fig.1(a).
\vspace{-5pt}
\subsection{Model Architecture}
\vspace{-5pt}
\subsubsection{The Latency-Controlled Streaming Encoder}
\vspace{-5pt}
%The latency-controlled streaming transformer encoder (LC-STE) is very similar to the encoder of the speech-transformer \cite{vaswani2017attention, dong2018speech},
The transformer encoder generally consists of a convolutional down-sampling block, a positional embedding, and $N$ transformer encoder blocks. The transformer encoder block is composed of a multi-head self-attention block and a feed-forward network layer \cite{vaswani2017attention}. The original transformer encoder depends on all context to compute the self-attention mechanism, which makes it powerful to model long-range temporal information, and prevents it from being applied to streaming speech recognition \cite{dong2018speech}. As depicted in Fig.1(b), we proposed a new streaming encoder named latency-controlled streaming transformer encoder (LC-STE), which improves the transformer encoder in two aspects.

% which includes a convolutional down-sampling block, a positional embedding and $N$ transformer encoder block. What's more, the transformer encoder is composed of multi-head self-attention block and feed-forward network layer. In order to make the encoder able to model streaming acoustic features, we improve the transformer encoder in two aspects.
% fixed-range previous context and ignore the future context  It's obviously that the future context will improve the performance,
% stacking many blocks make the model achieve better performance \cite{pham2019very}
On the one hand, we force the self-attention of LC-STE to focus on the fixed-range previous context and ignore the future context completely. It is well known that extending the range of future context will improve the performance of the model. However, with the increase of network depth, it will lead to excessive dependence on future information, making the latency of inference increase sharply. The streaming self-attention (SSA) can be expressed as follows.
\vspace{-5pt}
\begin{equation}
SSA(\bm{Q}_{t}, \bm{K}_{t-\tau:t}, \bm{V}_{t-\tau:t})=softmax(\frac{\bm{Q}_t\bm{K}_{t-\tau:t}^{\top}}{\sqrt{d_k}})\bm{V}_{t-\tau:t}
\end{equation}
where $\tau$ means the range of the previous context. $\bm{Q}_t$ is the query vector at the time $t$. $\bm{K}$ and $\bm{V}$ indicate the context. $d_k$ is the dimension of $\bm{K}$.

On the other hand, inspired by DeepSpeech2 \cite{amodei2016deep}, we put a latency-controlled context layer at the top of the encoder to model the fixed-range future context. The context layer is composed of a 1D convolution layer with kernel size $\varepsilon + 1$. The $\varepsilon$ means the range of future context. The acoustic encoded states can be expressed as $\hat{\bm{h}}_t$.
\vspace{-5pt}
\begin{equation}
\hat{\bm{h}}_t = \sum_{i=0}^{\varepsilon}\bm{w}_i\bm{h}_{t+i}+\bm{b}_i
\end{equation}
Where $\bm{h}$ is the output of the last streaming transformer encoder block. $\bm{w}$ and $\bm{b}$ indicate the weight and bias respectively. The latency can be expressed as $40 \times (\varepsilon +1)\text{ms}$, where $40\text{ms}$ means the convolutional down-sampling block can reduce the length of input sequences with frame shift 10ms by 4 times. With the latency-controlled context layer, the computation of ideal latency does not rely on the depth of the encoder.
\vspace{-5pt}
\subsubsection{The CTC Decoder and Transformer Decoder}
\vspace{-5pt}
The CTC decoder contains only a linear project layer, which is utilized to compute the label posterior probabilities of CTC. The transformer decoder is the same as it in speech transformer \cite{dong2018speech}, which is composed of masked multi-head self-attention, multi-head cross-attention, feed-forward network and positional embedding. During training, the decoder adopts a triangle-like mask to mask force the decoder to focus on the previous tokens \cite{vaswani2017attention}. We compute the CTC loss $\mathcal{L}_{CTC}$ and the cross entropy loss $\mathcal{L}_{CE}$ with label smooth for the CTC decoder and transformer decoder respectively. These three components are optimized jointly. The joint loss is expressed as $\mathcal{L}_{Joint}$.
\vspace{-5pt}
\begin{equation}
\mathcal{L}_{Joint}=\alpha\mathcal{L}_{CTC} + (1-\alpha) \mathcal{L}_{CE}
\end{equation}
where $\alpha$ is the weight of $\mathcal{L}_{CTC}$.
\vspace{-5pt}
% Because the CTC model cannot model the dependencies of the output, its performance is worse than other end-to-end model.
\subsection{Inference}
\vspace{-5pt}
We introduce the two-stage inference method for streaming inference, as shown in Fig.1(c). We assume that a large enough number of candidates generated by the CTC model at the first-stage inference can compensate for the lack of language modeling ability. Therefore, we figuratively rename the process that selecting the best sequence from numerous candidates as On-In-A-Hundred(OAH). The inference process can be split into two stages: pre-selection and one-step scoring. At the pre-selection stage, the CTC decoder can generate $N$ candidate sequences by prefix beam search \cite{hannun2014first} \footnote{The code of prefix beam search is available at https://github.com/\\PaddlePaddle/DeepSpeech/tree/develop/decoders}. The number of candidates is equal to the width of beams. At the one-step scoring stage, the transformer decoder scores all the candidates. Different from the language model rescoring, the one-step scoring relies on the corresponding acoustic encoded states to match the selected candidate with the c pronunciation. Due to the transformer decoder can perform parallel computation, the scoring can be finished in one-step, which improve the efficiency of inference. As long as the candidates are diverse enough, we might be able to pick out the best one. The score of a sentence of length $L$ can be expressed as

%As we all know, the CTC model has a weak ability to model the dependencies between the output tokens. From another perspective, the CTC can generate diverse candidates. We believe that we can pick out the best candidate from hundreds of candidates.
\vspace{-10pt}
\begin{equation}
S_{y_{1:L}} = OneStepScoring(y_{1:L}, \hat{\bm{h}}_{enc}) * 1 / L
\end{equation}
where $\hat{\bm{h}}_{enc}$ means the corresponding acoustic encoded states. We apply length normalization to the scoring process to prevent the model from tending to select the short candidate as the predicted sequence.

In addition, our model support two inference mode, streaming and non-streaming. When we adopt the CTC decoder as the leading role and the transformer decoder to score the candidates generated by the CTC decoder, the model can perform the streaming inference. The process is name one-in-a-hundred. From another viewpoint, when we make the transformer decoder as the leading role and the CTC decoder as the assistant, the transformer decoder can still model the whole context and decode in a non-steaming fashion \cite{watanabe2017hybrid}. The non-streaming decoding process starts with the beginning token \texttt{<S/E>}. At every step, we interpolate the scores of two decoders. The model will repeat the above process until the end-of-sentence token is predicted.
\vspace{-5pt}
\section{Related Works}
\vspace{-5pt}
% RNN-T
%The most similar to our work is the two-pass end-to-end model \cite{sainath2019two}, which combines the RNN-T and LAS decoder. Compared with the two-pass end-to-end model, we combine the CTC model and transformer. The CTC model is more efficient in training and inference. Therefore, we can generate hundreds of candidate sequences and evaluate them in a very short time. For the RNN-T model, it is very challenging to generate many candidates by beam search during inference, which will cost plenty of time and memory. As well known, the RNN-T is more powerful to model the language. Therefore, our motivation is to compensate the CTC models for the lack of language modeling ability by picking out the best sequence from the various candidates.
%
%Triggered attention model \cite{moritz2019triggered} is a kind of the hybrid CTC and attention model for streaming speech recognition, which utilizes the spike-like posterior probabilities generated by CTC to segment the encoded states. Our model can be regarded as an improved CTC model, which utilizes the attention-based decoder to evaluate the results of the CTC decoder. What's more, our model support two inference mode, streaming and non-streaming.
Our work can be regarded as a good extension and supplement of the previous two-pass end-to-end model \cite{sainath2019two}. The inference of these two methods can be divided into two stages, generating N-best candidates and selecting the best candidates by the extra decoder. However, there are at least three significant differences between these two works.

Firstly, the previous two-pass method is applied to the hybrid RNNT and LAS model. It utilizes the RNN-based LAS decoder to scoring the candidates, which is hard to perform in parallel. However, we introduce the two-pass method into the classical hybrid CTC and attention model and utilize a transformer decoder to rescore the candidate in one step, which is faster and more effective.

Secondly, it’s well known that the RNNT model has a strong ability to model language. For the hybrid RNNT and LAS model, the two-stage can be regarded as the fusion of two conditional language models. By contrast, the CTC model lacks the ability of modeling language. The second stage can be considered as a process that fuses linguistic information effectively.

Finally, it’s very challenging for RNNT to generate many candidate sequences by beam search during inference, which will cost plenty of time and memory. However, the CTC model performs more effectively during inference and training. Therefore, we try to generate up to one hundred candidates in the first stage of inference, which far exceeds the candidates from RNNT models. We assume that a large enough number and enough diversity of candidates can compensate the CTC model for the lack of language modeling ability.

\vspace{-5pt}
\section{Experiments and Results}
\vspace{-5pt}
\subsection{Dataset}
\vspace{-5pt}
In this work, all experiments are conducted on a public Mandarin speech corpus AISHELL-1\footnote{https://openslr.org/33/}.The training set contains
about 150 hours of speech (120,098 utterances) recorded by 340 speakers. The development set contains about 20 hours (14,326 utterances) recorded by 40 speakers. And about 10 hours (7,176 utterances / 36109 seconds) of speech is used as the test set. The speakers of different sets are not overlapped.
\vspace{-5pt}
\subsection{Experimental Setup}
\vspace{-5pt}
For all experiments, we use 40-dimensional FBANK features computed on a 25ms window with a 10ms shift. We choose 4233
characters (including a padding symbol \texttt{<PAD>}, an unknown symbol \texttt{<UNK>}, and an start-or-end-of-sentence symbol \texttt{<S/E>}) as modeling units. \texttt{<S/E>} is also utilized as a blank symbol for CTC Decoder.

Our model \footnote{Our model is built on the open-source code https://github.com/\\ZhengkunTian/OpenTransformer} consists of 12 encoder blocks and 6 decoder blocks. There are 4 heads in multi-head attention. The 2D convolution front end utilizes two-layer time-axis CNN with ReLU activation, stride size 2, channels 320, and kernel size 3. Both the output size of the multi-head attention and the feed-forward layers are 512. The hidden size of feed-forward layers is 768. The range of the self-attention in streaming transformer encoder blocks is limited from 10 frames on the left to the current position ($\tau=10$). We adopt an Adam optimizer with warmup steps 12000 and the learning rate scheduler reported in \cite{vaswani2017attention}. After 100 epochs, we average the parameters saved in the last 30 epochs. We also use the time mask and frequency mask method proposed in \cite{park2019specaugment} instead of speed perturbation.

We use the character error rate (CER) to evaluate the performance of different models. For evaluating the inference speed of different models, we decode utterances one by one to compute real-time factor (RTF) on the test set. The RTF is the time taken to decode one second of speech. All experiments are conducted on a GeForce GTX TITAN X 12G GPU.

\vspace{-5pt}
\subsection{Results}
\vspace{-5pt}
\subsubsection{Comparison of models with different CTC weights}
\vspace{-5pt}
We first compare the models with different CTC weights. We set the range of future context in latency-controlled context layer to 10 and adopt beam search with width 10 during inference. We evaluate the CER on the development and test set in two ways, which are one-pass beam-search (marked as \texttt{OPS}) and two-pass one-in-a-hundred (marked as \texttt{OAH}). This naming setting is still adopted in subsequent experiments.

As shown in Table.1, the model with CTC weight 0.1 can achieve the best performance on development and test test. Furthermore, it's obvious that applying our OAH strategy can generally improve the performance of the model. However, with the increase of CTC weight, the performance of the model gradually deteriorated. We assume that the weight $\alpha$ can balance the importance of the CTC model in the training process. An inappropriate weight will make the model unbalanced in the training process, leading to the decline of model performance. When the weight $\alpha$ is 1.0, the transformer decoder is discarded. Under this condition, the model can be regarded as a CTC model. By contrast, we find that the joint training with an appropriate weight can improve performance.

%\begin{table}[h]
%	\caption{Comparison of models with different CTC weights (CER \%).}
%	\centering
%	\label{tab:table1}
%	\begin{tabular}{|c|c|c|c|c|}
%		\hline
%		\multirow{2}{*}{CTC Weight $\alpha$} & \multicolumn{2}{c|}{Dev} & \multicolumn{2}{c|}{Test} \\ \cline{2-5}
%		& CTC     & OAH     & CTC        & OAH        \\ \hline
%		0.1                         &     \textbf{8.22}     &     \textbf{6.92}      &       \textbf{9.25}       &   \textbf{7.76}      \\ \hline
%		0.2                         &     8.26     &     7.04     &      9.36       &     8.02        \\ \hline
%		0.3                         &       8.31     &      7.14       &      9.67        &     8.18       \\ \hline
%		0.5                         &       8.89       &    7.55       &     10.21         &    8.69        \\ \hline
%		0.7                         &       9.53      &     8.03       &     11.03      &     9.26        \\ \hline
%		1.0                       &        8.88      &      -     &      10.27        &      -      \\ \hline
%	\end{tabular}
%\end{table}

\begin{table}[h]
	\caption{Comparison of models with different CTC weights (CER \%).}
	\vspace{-5pt}
	\centering
	\label{tab:table1}
	\begin{tabular}{|c|c|c|c|c|c|c|c|}
		\hline
		\multicolumn{2}{|c|}{Weight $\alpha$}	 & 0.1 & 0.2 & 0.3  & 0.5 & 0.7 & 1.0 \\
		\hline
		\hline
		\multirow{2}{*}{Dev} & OPS & \textbf{8.22} & 8.26 & 8.31 & 8.89 & 9.53 & 8.88 \\ \cline{2-8}
		& OAH & \textbf{6.92} & 7.04 & 7.14 & 7.55 & 8.03 & - \\
		\hline
		\hline
		\multirow{2}{*}{Test} & OPS & \textbf{9.25} & 9.36 & 9.67 & 10.21 & 11.03 & 10.27 \\ \cline{2-8}
		& OAH & \textbf{7.76} & 8.02 & 8.18 & 8.69 & 9.26  & - \\
		\hline
	\end{tabular}
	\vspace{-15pt}
\end{table}

\subsubsection{Comparison of models with different ranges of future context}
\vspace{-5pt}
We evaluate the models with different range $\varepsilon$ of future context in Table 2. We set the CTC weight of all models to 0.1 and adopt the beam search with width 10 during inference. The latency is proportional to the range of future context. It can be calculated by $40 \times (\varepsilon + 1)$ ms. It appears that the more future information the model focuses on, the better performance it achieves. Meanwhile, it will increase the computation and the real-time rate. When the range $\varepsilon$ is set to 10 and 20, there is no significant difference in the performance. Considering that the model that focuses on the next 10 frames can achieve lower latency and faster inference speed, we set the range of future context to 10 in subsequent experiments.

%\begin{table}[t]
%	\caption{Comparison of models with different ranges of future context (CER \%).}
%	\centering
%	\label{tab:table2}
%	\begin{tabular}{|c|c|c|c|c|c|c|c|}
%		\hline
%		\multirow{2}{*}{$\varepsilon$} & \multirow{2}{*}{Latency} & \multicolumn{2}{c|}{Dev} & \multicolumn{2}{c|}{Test} & \multicolumn{2}{c|}{RTF} \\ \cline{3-8}
%		&                          &       CTC      &     OAH    &     CTC      &      OAH     &     CTC      &      OAH   \\ \hline
%		0       &    40ms    & 11.70 & 8.75  &    13.15        &     10.18        &     0.0175       &  0.0271       \\ \hline
%		1       &    80ms    & 10.79 & 8.02  &    12.18        &     9.26        &      0.0209       &  0.0223          \\ \hline
%		3       &    160ms   &  9.27 & 7.35   &   10.39       &     8.36        &     0.0241       &  0.0271          \\ \hline
%		5       &    240ms   & 8.31 & 6.97    &   9.43        &    7.82        &      0.0254       &  0.0286          \\ \hline
%		10       &   440ms   & 8.21 & 6.91    &    \textbf{9.25}      &     7.76         &      0.0263       &  0.0281         \\ \hline
%		20       &   840ms   & \textbf{8.07} & \textbf{6.82}    &   9.26       &     \textbf{7.73}        &     0.0274      &  0.0295          \\ \hline
%
%	\end{tabular}
%	\vspace{-5pt}
%\end{table}

\begin{table}[h]
	\caption{Comparison of models with different ranges of future context (CER \%).}
	\vspace{-5pt}
	\centering
	\label{tab:table2}
	\begin{tabular}{|c|c|c|c|c|c|c|}
		\hline
		\multicolumn{2}{|c|}{$\varepsilon$}	 & 0 & 1 & 5 & 10 & 20 \\
		\hline
		\hline
		\multicolumn{2}{|c|}{Latency(ms)} & 40 & 80 & 240 & 440 & 840 \\
		\hline
		\hline
		\multirow{2}{*}{Dev} & OPS & 11.70 & 10.79 & 8.31 & 8.21 & \textbf{8.07} \\ \cline{2-7}
		& OAH & 8.75 & 8.02 & 6.97 & 6.91 & \textbf{6.82} \\
		\hline
		\hline
		\multirow{2}{*}{Test} & OPS & 13.15 & 12.18 & 9.43 & \textbf{9.25} & 9.26 \\ \cline{2-7}
		& OAH & 10.18 & 9.26 & 7.82 & 7.76 & \textbf{7.73} \\
		\hline
		\hline
		\multirow{2}{*}{RTF} & OPS & 0.0175 & 0.0209 & 0.0254 & 0.0263 & 0.0274 \\ \cline{2-7}
		& OAH & 0.0217 & 0.0223 &  0.0286 & 0.0281 & 0.0295 \\
		\hline
	\end{tabular}
	\vspace{-15pt}
\end{table}

%\begin{table}[t]
%	\caption{Comparison of models with different ranges of future context (CER \%).}
%	\centering
%	\label{tab:table2}
%	\begin{tabular}{|c|c|c|c|c|c|}
%		\hline
%		\multirow{2}{*}{$\varepsilon$} & \multirow{2}{*}{Latency} & \multicolumn{2}{c|}{Test} & \multicolumn{2}{c|}{RTF} \\ \cline{3-6}
%		&                          &       CTC      &     OAH    &     CTC      &      OAH      \\ \hline
%		0       &    40ms       &    13.15        &     10.18        &     0.0175       &  0.0271       \\ \hline
%		1       &    80ms       &    12.18        &     9.26        &      0.0209       &  0.0223          \\ \hline
%		2       &    120ms       &   10.39       &     8.36        &     0.0241       &  0.0271          \\ \hline
%		5       &    240ms       &   9.43        &    7.82        &      0.0254       &  0.0286          \\ \hline
%		10       &   440ms       &    \textbf{9.25}      &     7.76         &      0.0263       &  0.0281         \\ \hline
%		20       &   840ms       &   9.26       &     \textbf{7.73}        &     0.0274      &  0.0295          \\ \hline
%
%	\end{tabular}
%	\vspace{-5pt}
%\end{table}

\vspace{-5pt}
\subsubsection{Comparison of the model with different beam widths}
\vspace{-5pt}
We choose the model with the CTC weight 0.1 and the future context range 10 to conduct these experiments in Table 3. The results show that the beam widths have little effect on the performance of the CTC decode results (without OAH). We assume that the CTC part of our model has learned a very sharp posterior probability distribution, which leads to a numerous difference between the probabilities of search paths. However, the beam width plays a vital role in our two-stage inference method. The model with OAH scoring can achieve up to 20\% reduction in CER compared to the performance of the base CTC model (without OAH). The beam widths are equal to the number of candidate sequences. The larger the beam width, the better performance the model achieves. The model with beam width 50 achieves a CER of 7.41\% and an RTF of 0.0380. When the width is large than 50, the performance will decline. We assume that generating more hypotheses with larger beam sizes will also produce more wrong sequences with better score, which may have a negative impact on the final performance of the model \cite{zhou2020beam}.

\begin{table}[h]
	\caption{Comparison of the model with different beam widths (CER \%).}
	\vspace{-5pt}
	\centering
	\label{tab:table3}
	\begin{tabular}{|c|c|c|c|c|c|c|}
		\hline
		\multicolumn{2}{|c|}{Beam Width} & 1 & 5 & 10 & 50 & 100 \\
		\hline
		\hline
		\multirow{2}{*}{Dev} & OPS & 8.22  & 8.22  & 8.22  & 8.21 & 8.21 \\ \cline{2-7}
		& OAH & - & 7.15 & 6.92 & \textbf{6.73} & 6.76 \\
		\hline
		\hline
		\multirow{2}{*}{Test} & OPS & 9.25 & 9.25 & 9.25 & 9.25 & 9.25 \\ \cline{2-7}
		& OAH & - & 8.09 & 7.76 & \textbf{7.41} & 7.49 \\
		\hline
		\hline
		\multirow{2}{*}{RTF} & OPS & 0.0202 & 0.0216 & 0.0263 & 0.0262 & 0.0421 \\ \cline{2-7}
		& OAH & - & 0.0212 & 0.0281 & 0.0380 & 0.0487 \\
		\hline
	\end{tabular}
	\vspace{-10pt}
\end{table}

%\begin{table}[t]
%	\caption{Comparison with other models (CER \%).}
%	\centering
%	\label{tab:table4}
%	\begin{threeparttable}
%		\begin{tabular}{|l|c|c|c|}
%			\hline
%			Model & Is Streaming & Test & RTF \\ \hline
%			TDNN-Chain(Kaldi) \cite{povey2016purely}   &   N   &   7.45   &   - \\ \hline
%			Speech-Transformer* \cite{dong2018speech}  &   N    &    6.47    &   0.0564   \\ \hline
%			SA-Transducer \cite{Tian2019} &    N &  9.30  &   0.1536   \\ \hline
%			ST-NAT \cite{tian2020spike}  &    N &  7.67  &  0.0056   \\ \hline
%			Sync-Transformer \cite{tian2020synchronous}  &    Y &  8.91  &  0.1183   \\
%			\hline
%			\hline
%			SAN-CTC* \cite{salazar2019self} &      N  &  7.80   &  0.0269  \\ \hline
%			SAN-CTC*  &  Y  &  10.27  &  0.0248   \\
%			\quad+ RNNLM Rescoring  &    Y &  11.32    &   0.0385  \\
%			\quad+ TransLM Rescoring  &    Y &  12.99    &  0.0384  \\
%			\hline
%			\hline
%			Our Model (OAH)  &   Y &   7.41   &  0.0380   \\ \hline
%			Our Model (NS)  &   N &   7.05   & 0.0812 \\ \hline
%		\end{tabular}
%		\begin{tablenotes}
%			\item[*] These models are re-implemented by ourselves according to the papers, which has the same parameters configuration as our model.
%			%\item[$\dagger$] We supplement the RTF of our previous two models.
%		\end{tablenotes}
%	\end{threeparttable}
%	\vspace{-5pt}
%\end{table}

\begin{table}[h]
	\caption{Comparison with other models (CER \%).}
	\vspace{-5pt}
	\centering
	\label{tab:table4}
	\begin{threeparttable}
		\begin{tabular}{|l|c|c|c|}
			\hline
			Model & Is Streaming & Test & RTF \\ \hline\hline
			TDNN-Chain(Kaldi) \cite{povey2016purely}   &   N   &   7.45   &   - \\ \hline
			Speech-Transformer* \cite{dong2018speech}  &   N    &    6.47    &   0.0564   \\ \hline
			SA-Transducer \cite{Tian2019} &    N &  9.30  &   0.1536   \\ \hline
			ST-NAT \cite{tian2020spike}  &    N &  7.67  &  0.0056   \\ \hline
			Sync-Transformer \cite{tian2020synchronous}  &    Y &  8.91  &  0.1183   \\
			\hline
			\hline
			SAN-CTC* \cite{salazar2019self} &      N  &  7.80   &  0.0269  \\ \hline
			SAN-CTC*  &  Y  &  10.27  &  0.0248   \\
			\quad+ RNNLM Rescoring  &    Y &  9.02    &   0.0385  \\
			\quad+ TransLM Rescoring  &    Y &  8.93    &  0.0384  \\
			\hline
			\hline
			Our Model (OAH)  &   Y &   7.41   &  0.0380   \\ \hline
			Our Model (Non-Streaming)  &   N &   7.05   & 0.0812 \\ \hline
		\end{tabular}
		\begin{tablenotes}
			\item[*] These models are re-implemented by ourselves according to the papers, which has the same parameters configuration as our model.
			%\item[$\dagger$] We supplement the RTF of our previous two models.
		\end{tablenotes}
		\vspace{-10pt}
	\end{threeparttable}
	\vspace{-10pt}
\end{table}

\vspace{-5pt}
\subsubsection{Comparison with other models}
\vspace{-5pt}
We also compare our model with other models. As shown in Table 4, our model with OAH can achieve a comparable performance with TDNN-Chain model \cite{povey2016purely} and ST-NAT \cite{tian2020spike}. What's more, our model has a real-time factor of 0.0380, which exceeds SA-Transducer \cite{Tian2019}, Sync-Transformer\cite{tian2020synchronous}, and speech-transformer \cite{dong2018speech}.

Our model with OAH achieves better performance compared to the CTC with language rescoring. We assume that the model can't distinguish between two grammatical sentences that have similar pronunciation only depending on linguistic information. Therefore, the acoustic information still plays an import role in the second rescoring stage.

In addition, our model is able to decoding the output sequence in a non-streaming fashion (NS). Under this condition, our model achieves a CER of 7.05\%. Compared with the speech-transformer with the same parameters configuration, our model has a little performance degradation.

\vspace{-5pt}
\section{Conclusions}
\vspace{-5pt}
In this paper, we improve the hybrid CTC and attention model and introduce a two-stage inference method named one-in-a-hundred (OAH). Our proposed model consists of three components, a latency-controlled streaming transformer encoder, a CTC decoder, and a transformer decoder. The latency-controlled streaming transformer encoder can model the streaming input feature sequence in very low latency. The streaming inference process can be split into two stages: sampling and one-step scoring. At the first stage, the CTC decoder can generate up to a hundred candidate sequences quickly. At the second stage, the transformer decoder score all the candidate sequences based on the corresponding acoustic encoded states in one step.  We conduct the experiments on a public Chinese mandarin dataset AISHELL-1. The results show that our proposed method can achieve up to 20\% reduction in CER compared to the baseline CTC model, which also proves our assumption that the two-stage OAH inference can compensate the CTC model for the lack of language modeling ability. What's more, our model also can perform non-streaming decoding with a little performance degradation.
% In the future, we will improve the model in two aspects. On the one hand, we will explore how to generate numerous candidates in a faster fashion. On the other hand, we will try our best to unify the streaming and non-streaming models into this framework without any performance degradation.

\bibliographystyle{IEEEtran}

\bibliography{mybib}

\end{document}